\documentclass[%
reprint,
superscriptaddress,
prl,
 amsmath,amssymb,
 aps,
]{revtex4-2}
\usepackage{color}
\usepackage{gensymb}
\usepackage{graphicx}
\usepackage{dcolumn}
\usepackage{bm}
\usepackage{hyperref}
\usepackage{epstopdf}
\usepackage{float}
\usepackage{listings}
\usepackage{times,mathptm,bm}
\usepackage{xcolor}
\usepackage[export]{adjustbox}
\usepackage{mhchem}

\begin{document}
\title{Emergent Isotropic–Nematic Transition in 3D Semiflexible Active Polymers}

\author{Twan Hooijschuur}
\affiliation{Institute for Theoretical Physics, University of Amsterdam,
Science Park 904, 1098XH Amsterdam, The Netherlands.}%
\affiliation{Van der Waals-Zeeman Institute, Institute of Physics, University of Amsterdam, 1098XH Amsterdam, The Netherlands.}
\author{Ehsan Irani}
\affiliation{Institute for Theoretical Physics, Georg-August University of Göttingen, Friedrich-Hund Platz 1, 37077 Göttingen, Germany.}
\author{Antoine Deblais}%
\affiliation{Van der Waals-Zeeman Institute, Institute of Physics, University of Amsterdam, 1098XH Amsterdam, The Netherlands.}
\author{Sara Jabbari-Farouji}
\altaffiliation[Correspondence to: ]{\hyperlink{s.jabbarifarouji@uva.nl}{s.jabbarifarouji@uva.nl}}
\affiliation{Institute for Theoretical Physics, University of Amsterdam,
Science Park 904, 1098XH Amsterdam, The Netherlands.}%

\date{\today}

\begin{abstract}
Active semiflexible filament collectives, ranging from motor-driven cytoskeletal filaments to slender organisms such as cyanobacteria and worm aggregates, abound in nature. Yet how activity and flexibility jointly govern their organization, especially Isotropic-Nematic (I--N) transition, remains poorly understood. Performing large-scale Brownian dynamics simulations of 3D active semiflexible polymers with varying flexibility degrees, we show that tangential active forces systematically shift the I--N transition to higher densities, with the shift controlled by  the flexibility degree and activity strength. Strikingly, activity alters the nature of the transition: discontinuous at low strengths, continuous at moderate strengths, and ultimately suppressed at high activity levels.
The delayed I–N transition originates from enhanced collective bending fluctuations, resulting in chain shrinkage and enlargement of effective confinement tube. At moderate activity levels, these fluctuations can trigger large-scale excitations that stochastically drive temporal transitions between nematic and isotropic states, indicating an activity-induced instability of the nematic field. We summarize this behavior in non-equilibrium state diagrams of density and activity for different flexibility degrees. 
\end{abstract}

\maketitle

Active polymers and filaments~\cite{active_poly_review1,active_poly_review2}, self-driven flexible elongated entities, are ubiquitous across all scales from microscopic intracellular biopolymers  ~\cite{Keber2014,conboy2024cytoskeletal} and slender bacteria~\cite{mirabilis_Vibrio,Cyanobacteria_23} to macroscopic worms~\cite{Ozkan-Aydin2021,aquarium_worms}, snakes~\cite{Hu2009,Cohen2023}, and soft robotic grippers~\cite{Becker2022}. Active polymers and filaments represent a distinct class of active matter, where internal degrees of freedom give rise to rich dynamics, including reshaping, knotting~\cite{zhang2025_knot,Vatin_25} and topological entanglement \cite{hu2016entangled,deblais2023worm,aquarium_worms,tejedor2019reptation,Becker2022,ubertini2024universal,breoni2025giant}.
This interplay between flexibility and activity generates non-equilibrium behaviors absent in rigid active particles~\cite{Gompper2020,elgeti_physics_2015}. Prominent examples include coexisting ordered states~\cite{Huber2018}, spontaneous flows~\cite{sanchez2012spontaneous,Lemma2019}, cell-like migration~\cite{conboy2024cytoskeletal,Ozkan-Aydin2021}, collective ``blob'' formation~\cite{aquarium_worms,Patil2023} and activity-induced mechanical responses~\cite{mizuno2007nonequilibrium,koenderink2009active,alvarado2013molecular, worms_rheology,ozkan2021,kurjahn2024,conboy2025actin,breoni2025giant}. These collective behaviors underpin  vital functions of active filaments such as self-organized transport, programmable pattern formation, and adaptive navigation. 

A paradigmatic manifestation of activity in filament collectives is active nematics~\cite{marchetti_hydrodynamics_2013,Doostmohammadi}. It is known that stiff active filaments form a microscopically driven liquid crystal, where activity fuels the continual creation and annihilation of topological defects~\cite{marchetti_hydrodynamics_2013,Doostmohammadi,Wensink2012,Sanchez2012,Joshi2019,dry_active}. Unlike passive nematics, where orientational order is stabilized by free-volume effects and translational entropy gain~\cite{Onsager_49,Rods_IN}, high activity levels destabilize long-range alignment   giving rise to chaotic spatiotemporal dynamics \cite{Wensink2012,thampi2016active,alert2022active,Giomi2011}. While steady states of rigid active filaments are extensively studied~\cite{Wensink2012,Ginelli2010,IN_rods,dry_active,active_rods,Mandal2020,liu2025collective}, the influence of semiflexibility, especially in three dimensions (3D), remains largely unexplored. Existing 3D studies focus solely on the fully flexible limit~\cite{ubertini2024universal,breoni2025giant,liu2025collective,active_polymer_melt_25}, leaving open how activity and  finite flexibility degree together shape collective behavior and orientational order. Even in two dimensions, studies have mainly examined fixed-density semiflexible polymers under varying activity~\cite{Prathyusha2018,Duman2018,Joshi2019}, leaving the effect of activity on the density-driven isotropic–nematic (I--N) transition unresolved. Here, we address this gap by investigating the I--N transition of 3D active polar polymers across a range of flexibility degrees.

For passive polymers, chain flexibility is known to shift the (I--N) transition to higher densities as stiffness decreases~\cite{Binder_PRL_2016,Egorov2016}. This shift stems from anomalous nematic fluctuations driven by collective bending excitations within an effectively enlarged confinement tube, whose radius exceeds that expected from polymer concentration~\cite{Binder_PRL_2016,Egorov2016}. To probe how activity alters this behavior, we perform Brownian dynamics simulations of 3D tangentially-driven polymers~\cite{IseleHolder2015} with contour lengths exceeding their persistence lengths. We find that tangential active forces further shifts the I--N transition, with its magnitude set by polymer flexibility degree and activity level. Interestingly, the transition character changes from discontinuous at low activity levels, to continuous for moderate activity levels. In this continuous regime, we identify a new instability characterized by stochastic switching between homogeneous nematic states with long-range order and globally disordered states exhibiting local nematic order.

In our simulations, we model each active polymer as a linear bead-spring chain driven by tangential active forces~\cite{IseleHolder2015}, where the dynamics of each monomer follows overdamped Langevin dynamics. Adjacent beads along the chains interact with  harmonic spring potential with rest length $\ell_b$, while any pair of beads interact  with the short-ranged repulsive Weeks-Chandler-Andersen potential~\cite{WCA}. We choose the bond rest length to be equal to  the bead’s  diameter $\sigma$. Chains' stiffness is accounted by the bending potential of the form 
$U_{Bend}=\kappa (1-\cos\theta)$, where $\theta$ is the angle between two consequent bonds of a polymer and $\kappa$ is the bending stiffness constant setting the persistence length $\ell_p= \kappa \sigma/ k_BT$. The active force on each monomer is along the tangent of the polymer backbone, given by $\boldsymbol{F}^a =\tfrac{F^a}{2\ell_b}\,(\boldsymbol{b}_{i}^j+\boldsymbol{b}_{i+1}^j)$, where $\boldsymbol{b}_{i}^j$ refers to the $i$th bond vector of $j$th polymer, see the Supplementary Materials (Sup.) Sec.~I for further details ~\cite{SM}. We non-dimensionalize the system by choosing the monomer diameter $\sigma$ as the length scale, the thermal energy $k_B T$ as the energy scale, and $\tau = \gamma \sigma^2 / k_B T$ as the time scale, with the friction coefficient $\gamma$ set to unity.
 The system is thus governed by three key  parameters:  dimensionless monomer density $\rho \sigma^3$, active force $f^{a}=F^a\sigma/k_BT$ and bending stiffness $\kappa/k_BT$.

To investigate the collective behavior of active polymers, we simulate $M$ chains of length $N=32$ (contour length $L=31\sigma$) in a cubic box of size $L_{\rm box} = 64\sigma$. We consider four bending stiffness values, $\kappa/k_BT = 8,16,32,128,$ corresponding to  contour-to-persistence length ratios $0.24 < L/\ell_p < 3.88.$ The dimensionless active force is varied in the range $0 \le f^a \le 2,$ and the monomer density spans $0.1 \le \rho \sigma^3 \le 1.4,$ corresponding to volume fractions $\phi = \frac{\pi}{6} \rho \sigma^3$ between 0.05 and 0.73. The number of polymers ranges from $819 \le M \le 11{,}469,$ reaching up to $3.6 \times 10^5$ monomers in the densest systems. All simulations are performed with HOOMD-blue \cite{Anderson2020}, using an in-house implementation of tangential active forces. Systems are initialized from nematic configurations with polar active polymers randomly aligned along $\pm z$.
Stability is verified by monitoring the nematic order parameter over $10^4\tau$ ($\approx 2\tau_D$), where $\tau_D = \frac{NL^2}{6k_BT} \approx 5\times 10^3 \tau$ is the diffusion time of the center of mass of a passive polymer. Production runs extend over $20\tau_D,$ with averages taken over 20 independent configurations; in unstable regimes, simulations are extended up to $120\tau_D.$ From the saved trajectories during production runs, we compute all relevant observables, including global and local nematic order parameters associated with bond vectors.
 
\begin{figure}[t]
    \centering
    \includegraphics[width=1\linewidth]{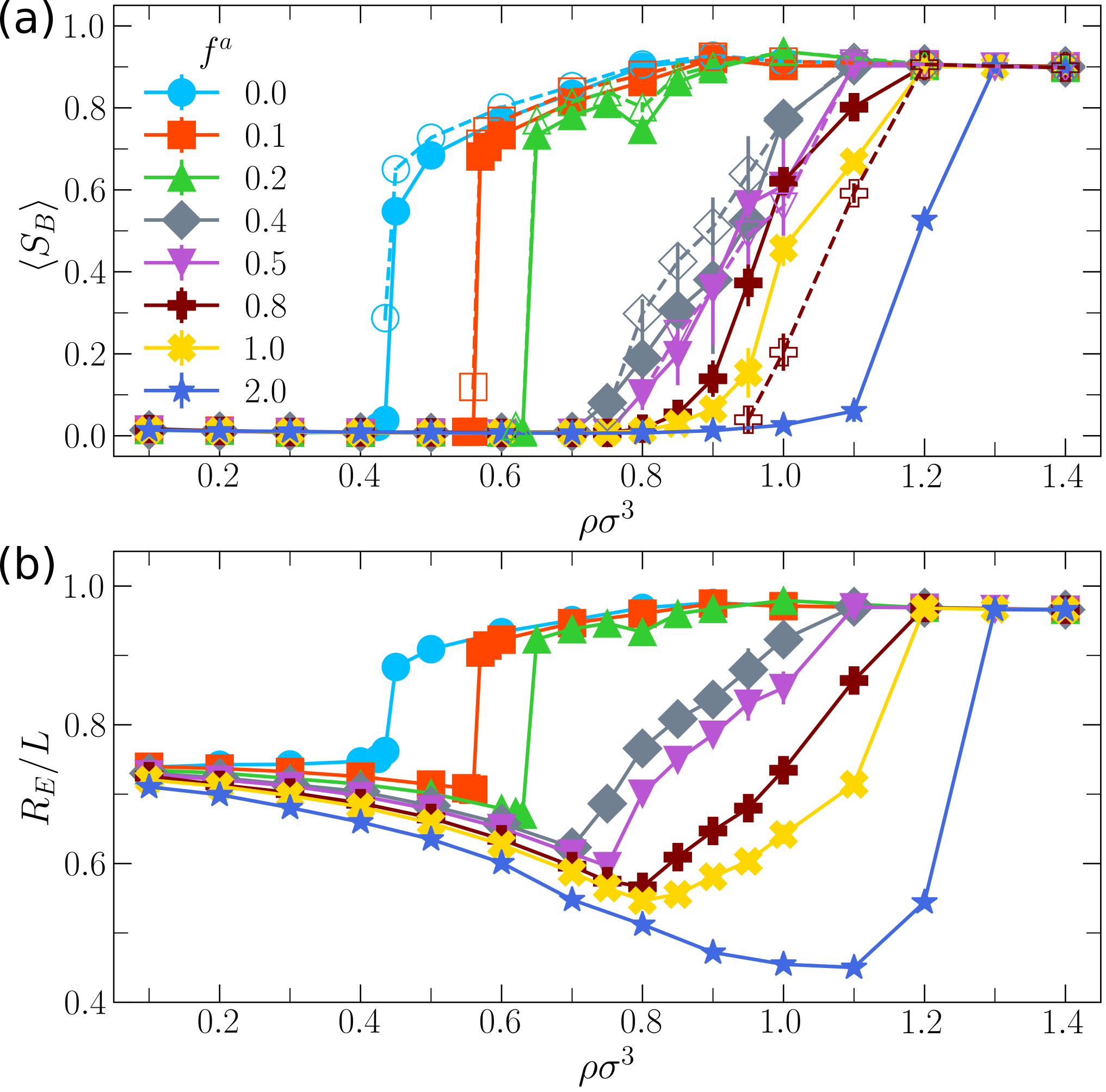}
    \caption{(a) Time-averaged global nematic order parameter of bond vectors, $\langle S_B \rangle$ (solid symbols) and (b) Mean end-to-end distance $R_E = \sqrt{\langle \mathbf{R}_e^2 \rangle}$ normalized by the contour length $L = 31\sigma$ as a function of density at fixed bending stiffness $\kappa/k_B T = 16$ for different active forces as shown in the legend. Open symbols in panel (a) (colored according to the same active-force code) represent estimates of $\langle S_B \rangle$ obtained from the mean end-to-end distance using $\langle S_B \rangle (R_E) = 3R_E/L - 2$.}

    \label{fig:Nematic_Order}
\end{figure}

To elucidate the interplay between activity  and flexibility on the I--N transition, we first examine orientational order parameters. Despite the polar nature of activity, we do not observe any global polar order of bond vectors even for the stiffest polymers and higher activities. The spatial correlation function of polar bond vectors displays a rapid, short-range decay, see Sup. Fig.~S1~\cite{SM}.
Next, we compute the global nematic order parameter $S_B$ from all bond vectors  $\boldsymbol{b}_i^j$ as the largest eigenvalue of $\mathbf{Q}=\frac{3}{2} [\frac{1}{M (N-1)}\sum_{i,j} \hat{\boldsymbol{b}_i^j}\otimes \hat{\boldsymbol{b}_i^j}  - \frac{1}{3} \mathbf{I}] $, see Sup.~Sec.~II.~A for details in \cite{SM}. Fig.~\ref{fig:Nematic_Order}(a) shows the time-averaged bond-order parameter $\langle S_B \rangle$ as a function of density for the different activity levels at a fixed flexibility degree $L/\ell_p \approx 1.94$ ($\kappa/k_BT = 16$). For passive semiflexible polymers ($f^a = 0$), $\langle S_B \rangle$ exhibits a sharp increase with density, confirming the first-order nature of the I--N transition, in agreement with prior studies~\cite{Binder_PRL_2016,Egorov2016}. This behavior is also reflected in the mean end-to-end distance $R_E = \sqrt{\langle \boldsymbol{R}_e^2 \rangle}$, which also increases sharply upon nematic alignment, see Fig.~\ref{fig:Nematic_Order}(b).
In the nematic regime, $\langle S_B \rangle$ is well described by $\langle S_B \rangle(R_E) = 3R_E/L - 2$ for $L/\ell_p \gg 1$~\cite{Egorov2016}, as indicated by the  open round symbols in Fig.~\ref{fig:Nematic_Order}(a). 

Upon introducing activity, two key effects emerge: (i) increasing $f^a$ shifts the I--N transition systematically to higher densities, and (ii) for $f^a > 0.2$, the sharp transition is replaced by a continuous one, indicating a fundamental change in the nature of the ordering process.

At low activity levels ($f^a \le 0.2$), the I–N transition shifts to higher densities while retaining its discontinuous character. In the nematic phase, both $\langle S_B \rangle$ and $R_E$ remain close to their passive values and converge at high densities, whereas in the isotropic regime, $R_E$ of active polymers decreases with increasing density, unlike their passive counterparts. In the dilute limit, $R_E$ and $\ell_p$ of active chains recover their passive values, indicating that the shrinkage originates from the interplay between activity and crowding (see Sup.~Fig.~S2~\cite{SM}). Frequent polymer collisions enhance bending fluctuations, effectively reducing $R_E$. Despite the delayed transition, weakly active polymers  follow the relation $\langle S_B \rangle(R_E) = 3R_E/L - 2$ once in the nematic regime.

At higher activity levels ($f^{a} > 0.2$), the sharp jump in $\langle S_B \rangle$ with density is replaced by a smooth, continuous increase, accompanied by strong temporal fluctuations, as evidenced by the error bars of the time-averaged nematic order. Concurrently, the mean end-to-end distance $R_E$ exhibits a pronounced shrinkage with increasing density prior to the transition due to activity-induced collisions, whereas it increases smoothly once nematic order sets in ($\langle S_B \rangle > 0.1$), remaining insensitive to nematic order fluctuations. For moderate activity levels ($f^{a} = 0.4$–$0.5$), the data still approximately follow the relation $\langle S_B \rangle(R_E) = 3R_E/L - 2$ at finite $\langle S_B \rangle$, but this correspondence breaks down at higher activity levels ($f^{a} \gtrsim 0.8$), where nematic order decouples from chain extension.

\begin{figure}[t]
    \centering
    \includegraphics[width=1\linewidth]{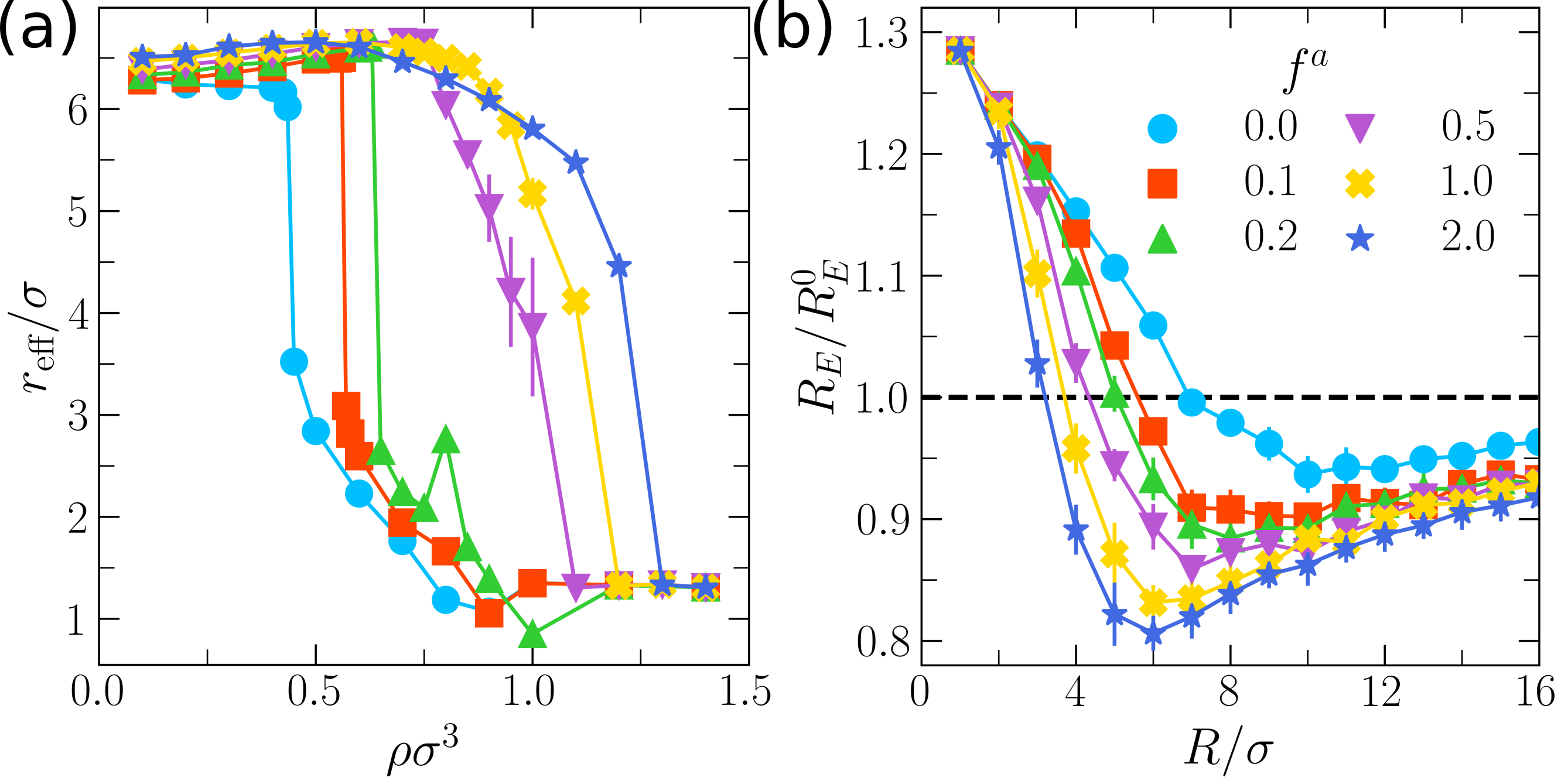}
    \caption{(a) Radius of effective confinement tube $r_{\text{eff}}$ for $\kappa/k_BT = 16$ as a function of density for varying active forces.
(b) Mean end-to-end distance $R_E$ of a tangentially driven polymer confined in a cylindrical channel of radius $R$, normalized by the passive unconfined single-chain value $R_E^0$. The dashed line indicates $R_E/R_E^0 = 1$, and its intersections with $R_E/R_E^0$ curves mark the crossover from compressed to stretched conformations under confinement.
}
\label{fig:Cylinder_Reff}
\end{figure}

Now, we show that the activity-induced chain shrinkage prior to the I–N transition,  driving its shift to higher densities, can be understood in terms of  confinement of an active chain  within an effective tube of its neighbors.
Following the framework for passive semiflexible polymers~\cite{Binder_PRL_2016,Egorov2016}, we extract an effective tube radius $r_{\text{eff}}$ from transverse backbone fluctuations relative to the end-to-end vector, see Sup.~Sec.~IV~\cite{SM}. Figure~\ref{fig:Cylinder_Reff}(a) shows $r_{\text{eff}}$ for $\kappa/k_BT=16$ across $ 0\le f^{a}\le 2$. It remains nearly constant in the isotropic regime, largely independent of activity, but drops after the I--N transition, signaling the onset of a strong confinement regime that leads to polymer stretching. Akin to the nematic order parameter, $r_{\text{eff}}$ decreases sharply at low activity levels, but varies more smoothly for larger active forces, ultimately reaching $r_{\text{eff}}\sim\sigma$ at high densities, enforcing stretched conformations, seen in Fig.~\ref{fig:Nematic_Order} (see also Sup.~Fig.~S3(a) in \cite{SM}).

In the isotropic regime, $r_{\text{eff}} \approx 6.5\sigma$, is significantly larger than the geometric estimate $r_\rho=\sqrt{N/(\pi \rho R_E)}$ deduced from the monomer density $\rho$ in a cylindrical tube of height $R_E$; $0.5<r_\rho\leq 2$ for $0.1<\rho\leq 1.4$ pointing to the collective chain deflections~\cite{Binder_PRL_2016}, see also Sup.~Fig.~S3(b) in \cite{SM}. The delayed drop in $r_{\text{eff}}$ reveals the origin of the shifted I–N transition: activity enhances collective bending fluctuations, sustaining an enlarged effective tube diameter up to higher densities. 

To directly test this confinement picture, we compare the mean chain conformations in collectives [Fig.~\ref{fig:Nematic_Order}(b)] with those of polymers confined in cylindrical tubes of radius $R \leq \ell_p$ with repulsive walls, see Sup.~Sec.~V in \cite{SM} for simulation details and Supplementary Videos 1 and 2. To quantify the coupling between confinement and polymer conformation, we measure the mean end-to-end distance $R_E$, normalized by its dilute-limit value $R_E^0$, as a function of tube radius $R$ as presented in Fig.~\ref{fig:Cylinder_Reff}(b). For wide tubes, $R_E<R_E^0$ due to wall interactions, whereas strong confinement (small $R$) enforces stretching with $R_E>R_E^0$. Increasing activity enhances the shrinkage and shifts the crossover to extended conformations toward smaller $R$, demonstrating that more active semiflexible polymers require stronger confinement to stretch. This explains the delayed I–N transition: at equal densities, activity enhances conformational fluctuations resulting in chain shrinkage, postponing the confinement-induced stretching that drives nematic order in passive systems.
 
\begin{figure}[t!]
  \centering
    \includegraphics[width=0.95\linewidth]{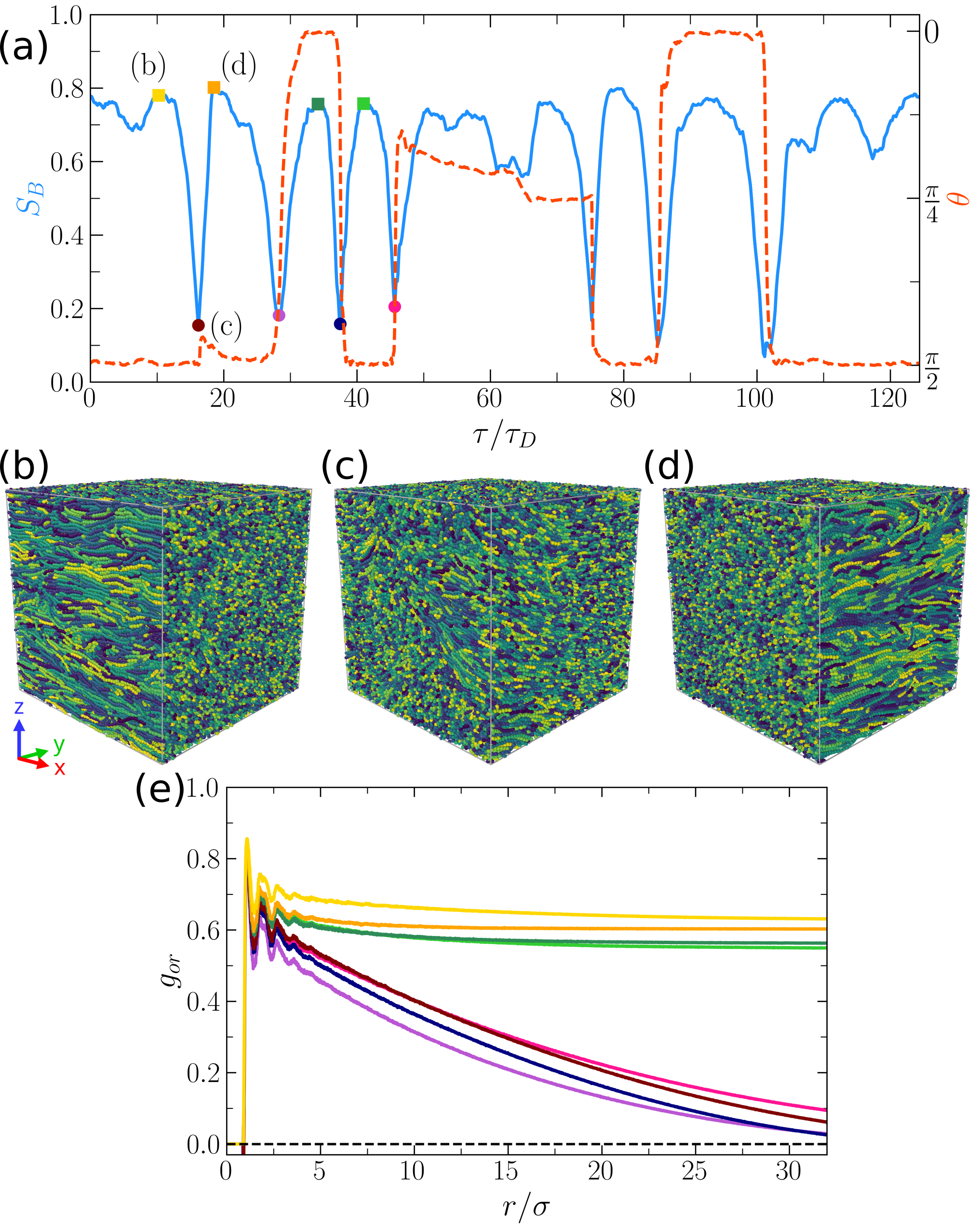}
  \caption{(a) Instantaneous global nematic order parameter $S_B$ (blue) and director angle $\theta$ (red) for $\kappa/k_BT = 16$, $\rho\sigma^{3} = 1.0$, and $f^{a} = 0.5$ as functions of time. The angle $\theta$ is measured relative to the $z$-axis, with $\theta = 0$ indicating alignment along $z$ and $\theta = \pi/2$ alignment in the $x$–$y$ plane. Time is normalized by the passive polymer center-of-mass diffusion time $\tau_D = \gamma N L^2 / 6k_B T$. Snapshots in panels (b)–(d) correspond to the black markers in (a), illustrating (b) alignment along $x$, (c) loss of global order due to instabilities, and (d) realignment along $y$. (e) Orientational pair correlation function $g_{\mathrm{or}}(r)$ of bond vectors at selected times marked by squares of the same color in panel (a).}
\label{fig:Osc}
\end{figure}  

At high-activity regime ($f^{a} \geq 0.5$), where the I--N transition becomes continuous, the instantaneous $S_B$ shows pronounced temporal fluctuations, signaling an instability of orientational order.
Fig.~\ref{fig:Osc}(a) shows an example at $f^{a}=0.5$ and $\rho\sigma^{3}=1.0$, where $S_B$  alternates between  nematic and disordered states on timescales of $\sim 10\tau_D$. The evolution of the director angle $\theta$ confirms that the large-amplitude fluctuations in the nematic order arise from large scale reorientation of bond vectors. Snapshots in Figs.~\ref{fig:Osc}(b)--(d) illustrate the sequence of events underlying these oscillations. At $t \approx 10\,\tau_D$, the polymers are globally aligned along the $x$-axis. As activity drives enhanced bending fluctuations, this order gradually destabilizes until, around $t \approx 16\,\tau_D$, the nematic alignment is fully lost during collective reorientation. The system then regains order, now oriented along the $y$-axis. The full temporal evolution of these structural rearrangements is shown in Sup.~Video~3.
As can be seen from Figs.~\ref{fig:Osc}(c), even when global order vanishes, local alignment persists. To quantify it, we compute the orientational pair correlation of bond vectors,
$g_{or}(r) = \langle P_2(\cos \alpha(r)) \rangle$, where $\alpha$ is the angle between bonds at separation $r$ at different moments as presented in Fig.~\ref{fig:Osc}(e). When $S_B$ is large, $g_{or}(r)$ plateaus, reflecting long-range order; when $S_B$ is small, it decays to medium-range order over half the box length. Analyzing energy densities of bend and splay deformations, we find that the I--N transition is marked by  dominance of bend over splay fluctuations (Sup.~Fig.~S4~\cite{SM}), and activity shifts the onset of this 
predominance to higher densities. 

\begin{figure}[t]
  \centering
\includegraphics[width=0.95\linewidth]{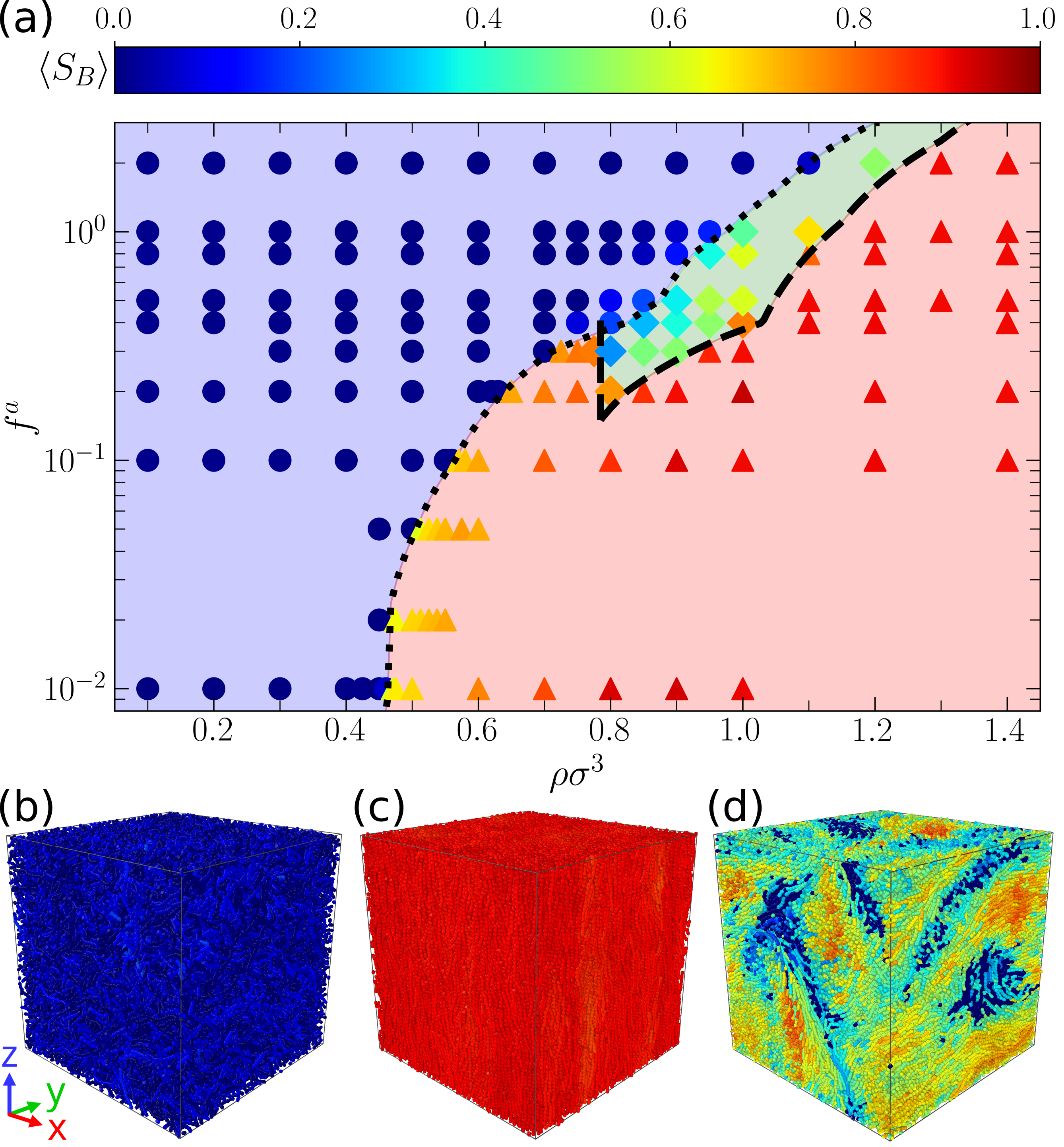}
  \caption{(a) State diagram for $\kappa/k_BT=16$ as a function of density and activity, with isotropic (disks), nematic (triangles), and unstable (diamond) regimes. Colors denote the time-averaged global nematic order parameter $\langle S_B \rangle$. Representative snapshots of (b) isotropic ($\rho\sigma^{3}=0.5$, $f^{a}=0.1$), (c) nematic ($\rho\sigma^{3}=0.8$, $f^{a}=0.1$), and (d) unstable ($\rho\sigma^{3}=0.9$, $f^{a}=0.5$) states, where polymers are colored by their local nematic order $S_B^{\text{loc}}$ using the same color code as panel (a).}
     \label{fig:Phase_Diagram}
\end{figure}

\begin{figure}[t]
    \centering
    \includegraphics[width=0.95\linewidth]{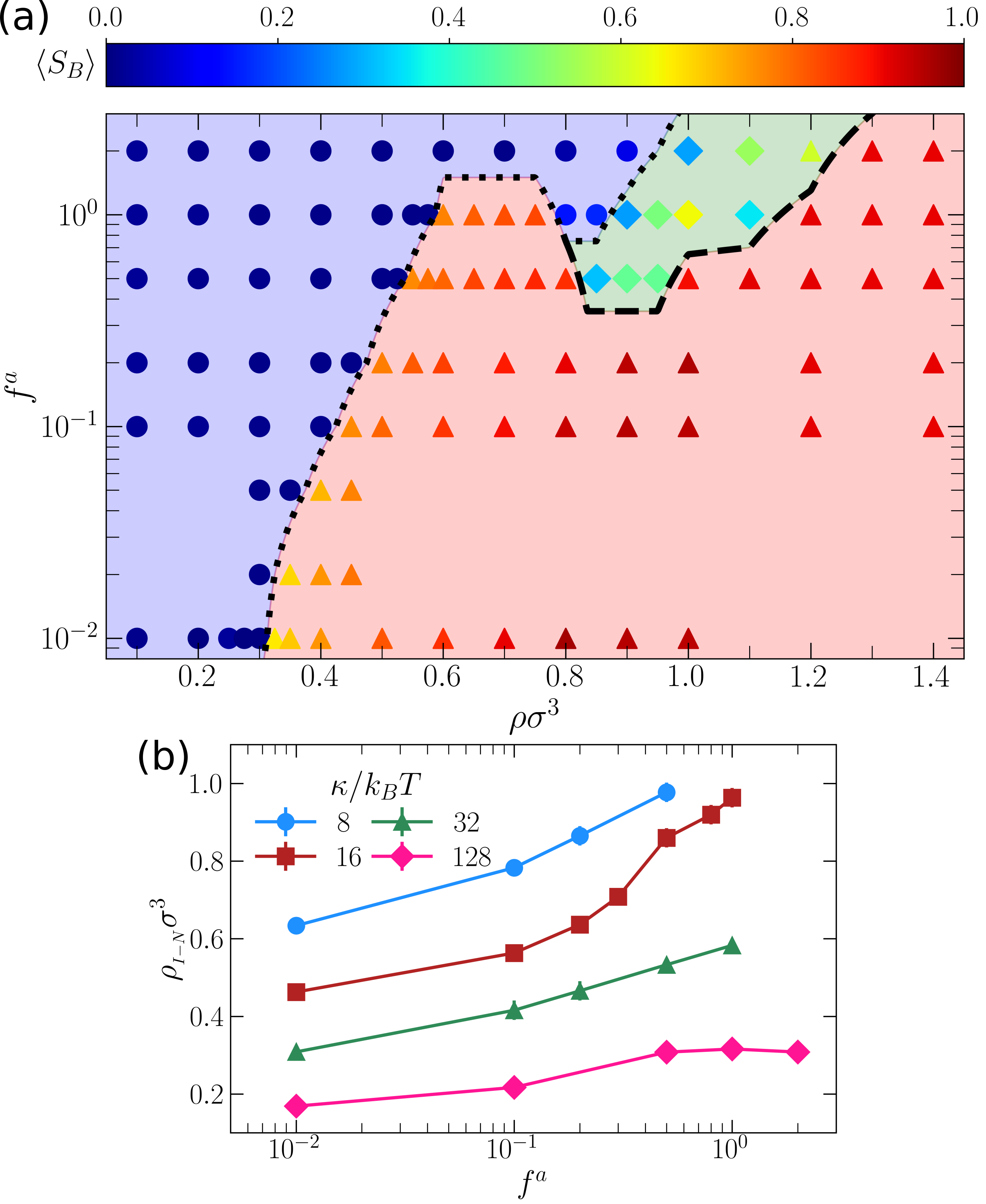}
    \caption{(a) State diagram for larger bending stiffness $\kappa/k_BT=32$, covering the same range of activities and densities as in  Fig.~\ref{fig:Phase_Diagram}(a) with identical symbol representation. (b) The I--N transition density $\rho_{\text{I--N}}$ as a function of active force for different bending stiffness values. }
    \label{fig:Kappa}
\end{figure}

We summarize all the observed regimes in the state diagram of Fig.~\ref{fig:Phase_Diagram}(a), shown as a function of density and activity, where each point is colored by the time-averaged global nematic order $\langle S_B \rangle$. At low activity levels, the isotropic and nematic phases are clearly separated, with the I--N boundary shifting continuously to higher densities as activity increases. The instability regime is identified by two criteria: (i) finite time-averaged global nematic order $\langle S_B \rangle > 0.2$ and (ii), $\sigma_{S_B}>0.01$, reflecting sufficiently large temporal fluctuations similar to the example of Fig.~\ref{fig:Osc}. This regime requires both sufficiently large activity and density. Representative snapshots of the different states are shown in Figs.~\ref{fig:Phase_Diagram}(b)–(d), where polymers are colored by their local nematic order $S_B^{\text{loc}}$, see Sup.~Sect.~II.B in~\cite{SM} for definition. In the isotropic example in Fig.~\ref{fig:Phase_Diagram}(b) ($\rho\sigma^{3}=0.5$, $f^a=0.1$) polymers show no orientational order, whereas in the nematic example in Fig.~\ref{fig:Phase_Diagram}(c)($\rho\sigma^{3}=0.8$, $f^{a}=0.1$) they are strongly aligned. In the unstable case shown in Fig.~\ref{fig:Phase_Diagram}(d) ($\rho\sigma^{3}=0.9$, $f^{a}=0.5$), the system exhibits coexisting regions of low and high local nematic order, exhibiting strong temporal fluctuations see Sup.~Video~4.

Finally, we examine the influence of bending stiffness $\kappa$ on the I--N transition. Figure~\ref{fig:Kappa}(a) shows the state diagram for $\kappa/k_BT = 32$ over the same range of activities and densities as Fig.~\ref{fig:Phase_Diagram}(a). At low activity levels, the I--N boundary follows that of the $\kappa/k_BT = 16$ case but extends up to $f^{a}=1.0$. An instability regime re-emerges at higher activity and density ($\rho\sigma^3 \gtrsim 0.8$). In contrast to the $\kappa/k_BT = 16$ case, where $\langle S_B \rangle$ increases with density within the instability regime, here it decreases, suggesting that enhanced bending fluctuations at higher stiffness hinder the establishment of global nematic order. At higher densities nematic order is restored. To examine the combined influence of flexibility and activity, we plot the I–N transition density, defined by $\langle S_B \rangle = 0.3$, as a function of $f^{a}$ for $\kappa/k_BT = 8, 16, 32,$ and $128$ in Fig.~\ref{fig:Kappa}(b). For a given $f^a$, increasing $\kappa$ shifts the transition to lower densities, consistent with passive-system results~\cite{Binder_PRL_2016,Egorov2016}; see Sup.~Figs.~S5~\cite{SM} for the full dependence of $\langle S_B \rangle$ and $R_E$ on $\kappa$ and $f^{a}$. Increasing stiffness also shifts the onset of the instability regime to higher activity and density values. 

In summary, we present the first systematic study of how activity affects the isotropic–nematic transition of semiflexible polymers in 3D. Activity profoundly alters the character of transition: at low activity levels, it shifts to higher densities while remaining discontinuous, but at higher activity levels it becomes continuous due to nematic-field instabilities that generate large-scale orientational fluctuations. Bending stiffness further modulates both the location of the transition and the extent of the resulting non-equilibrium steady states. Together, these findings uncover the intricate interplay between activity, density, and flexibility, establishing a foundation for understanding the mechanics, collective behavior, and self-organization of active polymer-like matter. Unlike earlier 2D studies at fixed density values~\cite{Henkes2011,Duman2018,Joshi2019}, our results reveal that activity-driven changes in the I–N transition nature, suggesting that this change  may be a generic feature of active polymers, meriting further exploration. Our findings call for new theories of active nematics for semiflexible polymers and motivate future work across active matter systems to determine how activity reshapes phase transitions out of equilibrium.

\begin{acknowledgements}
We acknowledge  L. Giomi, K.  Kruse, F. Toschi and  J. Yeomans for fruitful discussions. The computations were carried out on the Dutch National e-Infrastructure with the support of the SURF Cooperative.
\end{acknowledgements}

\newpage

\makeatletter
\def\thesection{\Roman{section}}
\def\thesubsection{\Alph{subsection}}
\makeatother
\clearpage
\onecolumngrid

\begin{center}
    \textbf{\large Supplementary Materials for:\\ Emergent Isotropic-Nematic Transition of 3D Active Semiflexible Filaments}
\end{center}
\bigskip

\setcounter{figure}{0}
\setcounter{table}{0}
\setcounter{equation}{0}
\renewcommand{\thefigure}{S\arabic{figure}}
\renewcommand{\thetable}{S\arabic{table}}
\renewcommand{\theequation}{S\arabic{equation}}

\renewcommand{\figurename}{SUP.~FIG.}






\section{I.~ Simulation method and details}\label{sec:SimMethod}
We model semiflexible active polymers as bead-spring chains consisting of $N$ monomers which are connected by harmonic springs. The position of the $i$th monomer with $ 1\le i\le N$ in the polymer labeled as $ 1\le j \le M$ is given by $\boldsymbol{r}_i^j$. The bond vectors within the  $j$th polymer connecting    $i$ and $i+1$th neighbor monomers  are defined as $\vec{b}_{i} ^j=\vec{r}_{i+1}^j-\vec{r}_{i}^j$.
 
 The dynamics of each monomer is described by the overdamped Langevin equation of motion:
\begin{equation}
    m\boldsymbol{\ddot{r}}_i^j=-\gamma \boldsymbol{\dot{r}}_i^j+\boldsymbol{F}^{a,j}_i-\sum_{k,l}\nabla_{\boldsymbol{r}_i^j}U(|\boldsymbol{r}_{i}^j-\boldsymbol{r}_{k}^l|)+\boldsymbol{F}^{r,j}_i,
\end{equation}
In this equation the mass of each monomer is given by $m$ and $\gamma$ describes the friction coefficient of the drag term chosen such that $m/\gamma  \ll 1$. All the conservative interactions between the monomers are included in the total potential energy $U_{tot}$, which consist of contributions of harmonic bond, the bending potential and the excluded volume interactions to prevent the crossing of polymers.    The harmonic potential between two adjacent monomers is given by
\begin{equation}\label{eq:U_Harmonic}
     U_{Bond}(r)=\frac{1}{2}k(r-\ell_b)^2,
\end{equation}
where $r$ is the distance between two consequent monomers in the same chain, $\ell_b$ is the equilibrium bond length and $k$ is the spring constant. 
To account for the degree of flexibility of the chain we incorporate a bending potential of the form
\[
    U_{Bend}=\kappa\big(1-\cos\theta_{i}^j\big),
\]
where $\theta_{i}^j$ is the angle between consecutive bonds as $\cos\theta_{i}^j=\hat{\textbf{b}}^j_{i}\cdot \hat{\textbf{b}}^j_{i+1}$ with $\hat{\textbf{b}}_{i}^j=\textbf{r}_{i,i+1}^j/r_{i,i+1}^j$. The bending stiffness $\kappa$ is related to the thermal persistence length as $\ell_p=\kappa\sigma/k_BT$. 

The excluded volume interactions between any two monomers a distance $r$ apart both within the same chain or in different polymers are modeled by the truncated and shifted Lennard Jones potential known as Weeks-Chandler-Anderson(WCA) potential ~\cite{WCA}.
\begin{equation}\label{eq:U_WCA}
    U_{WCA}(r)=\begin{cases}
    4\varepsilon\Big(\big(\frac{\sigma}{r}\big)^{12}-\big(\frac{\sigma}{r}\big)^6+\frac{1}{4}\Big), & \text{if $r\leq2^{1/6}\sigma$}.\\
    0, & \text{$r>2^{1/6}\sigma$}.
  \end{cases}
\end{equation}

 The active force $\boldsymbol{F}^{a,j}_i$ is tangential to the polymer backbone, following the model of Ref.~\cite{IseleHolder2015}, and can be written as
\begin{equation}
    \boldsymbol{F}^{a,j}_i =\frac{F^a}{2\ell_b}\big(\vec{b}_{i-1} ^j+\vec{b}_{i} ^j\big),
\end{equation}
where $\ell_b$ is the average bond distance and $\textbf{r}_{i,k}^j$ is the bond vector connecting monomers $i$ and $k$ on the $j$th polymer. For the first and last monomer, the forces are $\boldsymbol{F}_1^{a,j}=\frac{F^a}{2\ell_b}\vec{b}_{1} ^j$ and $\boldsymbol{F}_N^{a,j}=\frac{F^a}{2\ell_b}\vec{b}_{N-1} ^j$, respectively. Random thermal forces $\boldsymbol{F}^{r,j}_i$ are modeled as Gaussian white noise with $\langle\boldsymbol{F}^{r,j}_i\rangle=0$ and correlations $\langle\boldsymbol{F}^{r,p}_i(t)\boldsymbol{F}^{r,q}_j(t^\prime)\rangle=6k_BT\gamma\delta_{ij}\delta_{pq}\delta(t-t^\prime)$.  

  Our Brownian dynamics simulations were performed using the HOOMD-blue software package~\cite{Anderson2020},  with an in-house extension to apply tangential active forces to the polymers. HOOMD-blue was chosen due to it being optimized for large-scale MD simulations on GPU computational clusters.

\subsection{\label{subsec:units_sim_parameters}A.~ Units and simulation parameters}
 
To perform the simulations, we render the equations dimensionless. We choose the diameter of the monomer $\sigma$ as the unit of length $L_u=\sigma$, the strength of the WCA potential as the unit of energy $E_u=\varepsilon$,  $t_u=\tau=\gamma\sigma^2/\varepsilon$ as the unit of time, and the friction coefficient $\gamma $ is set to unity. These choices lead to the definition of dimensionless mass as $m^*=\frac{m \epsilon}{(\sigma \gamma)^2}$ and dimensionless active force $f^a=F^a \sigma/ k_BT $, which can also be interpreted as the monomeric Péclet number $Pe_m$~\cite{Bianco2018,IseleHolder2015}. As all polymers in our simulation have the same length $N=32$.
To ensure that we stay in  the overdamped limit $m/\gamma\ll 1$ and to avoid inertial effects~\cite{FazelzadehInertia2022},  we choose $m^*=0.01$ and the bond rest length $\ell_b=1$. The timestep is $\delta t=0.001\tau$ is chosen sufficiently  small to allow for simulations with high active forces. We set the temperature to unity $k_bT=1.0\varepsilon$. The spring constant of the bond potential is set to $k=5000\varepsilon/\sigma^2$, which is sufficiently stiff to ensure that the average bond distance is unchanged at high activity levels.
Activity is expressed through the dimensionless active force $f^{a}=F^a\sigma/k_BT$, which compares the diffusion time of a single monomer $\tau_{m}=\gamma/k_BT$ with the advection time $\tau_A=\gamma/F^a$. 

\section{II.~ Definitions and computation of observables}
\subsection{A.~ Global nematic order parameter of bond vectors}

In the main text, we characterized the I--N transition by the global nematic order parameter of bond vectors denoted by $S_B$. This is obtained from the nematic tensor of unit bond vectors as
\begin{equation}\label{eq:Q_Nem}
    \boldsymbol{Q}=\frac{3}{2}\Bigg(\frac{1}{M(N-1)}\sum_{i,j}^{N-1,M}\hat{\boldsymbol{b}}^{j}_i\otimes \hat{\boldsymbol{b}}^{j}_i-\frac{1}{3}\boldsymbol{I}\Bigg).
\end{equation}
Here, $\otimes$ is the tensor product over all the Cartesian coordinates, $\hat{\boldsymbol{b}}^{j}_i$ represents the unit bond vector of the $i$th bond of the $j$th polymer, $M$ is the total number of polymers, $N$ is the number of monomers per polymer, and $\boldsymbol{I}$ is the identity matrix. The nematic tensor $\boldsymbol{Q}$ yields three eigenvectors $\hat{\boldsymbol{v}}_i$ with corresponding eigenvalues $\lambda_i$, satisfying $\sum_i \lambda_i=0$ due to it being a traceless matrix. The global nematic order parameter is defined as the largest eigenvalue,
\[
    S_B=\max\{\lambda_i\},
\]
and the corresponding eigenvector provides the nematic director $\hat{\boldsymbol{n}}$. 

\subsection{B.~ Polymer-based local nematic order parameter}\label{sec:LocalNem}

At low activity  levels, the global order parameter fully characterizes the system. However, in the instability regime, a more local measure is needed. Large-scale fluctuations of the global nematic order lead to strong temporal inhomogeneities, and thus characterization of local nematic alignment is desirable.  

Hence, we calculate a local nematic order parameter in the vicinity of each polymer. To this end, we first identify neighboring bond vectors $\hat{\boldsymbol{b}} ^{i,m}$ of  polymer $m$ as those closer than the cutoff radius $r_c=1.5\sigma$ to any monomer of it and we add the them to the local cluster $c^{m}$.  From this collection of neighboring bonds, we construct a local nematic tensor $\boldsymbol{Q}^{m,Loc}$ and define a local nematic order parameter $S^{m,Loc}$.
 From this chain specific cluster of neighboring bonds $c^{m}$, we define a nematic tensor,
\begin{equation}\label{eq:Q_Nem_Local}
  \boldsymbol{Q}^{m,Loc}=\frac{3}{2}\Bigg(\frac{1}{N_c^{m}}\sum_{i\in c^m}^{N_c^m} \hat{\boldsymbol{b}}^{i,m}\otimes \hat{\boldsymbol{b}}^{i,m}-\frac{1}{3}\mathbb{I}\Bigg).
\end{equation}
Here is $N_c^{m}$ the total number of bonds in cluster $c^m$.  From this nematic tensor we can define a local nematic order parameter for every polymer $S^{m,Loc}$ as its largest eigenvalue. 
This order parameter allows us to visualize the spatial nematic structure of the system in different regimes, as shown in Fig.~4(b)-(d) of the main text.

\subsection{C.~ Spatial orientational correlation of polar order}
\begin{figure}[h]
    \centering
    \includegraphics[width=0.8\linewidth]{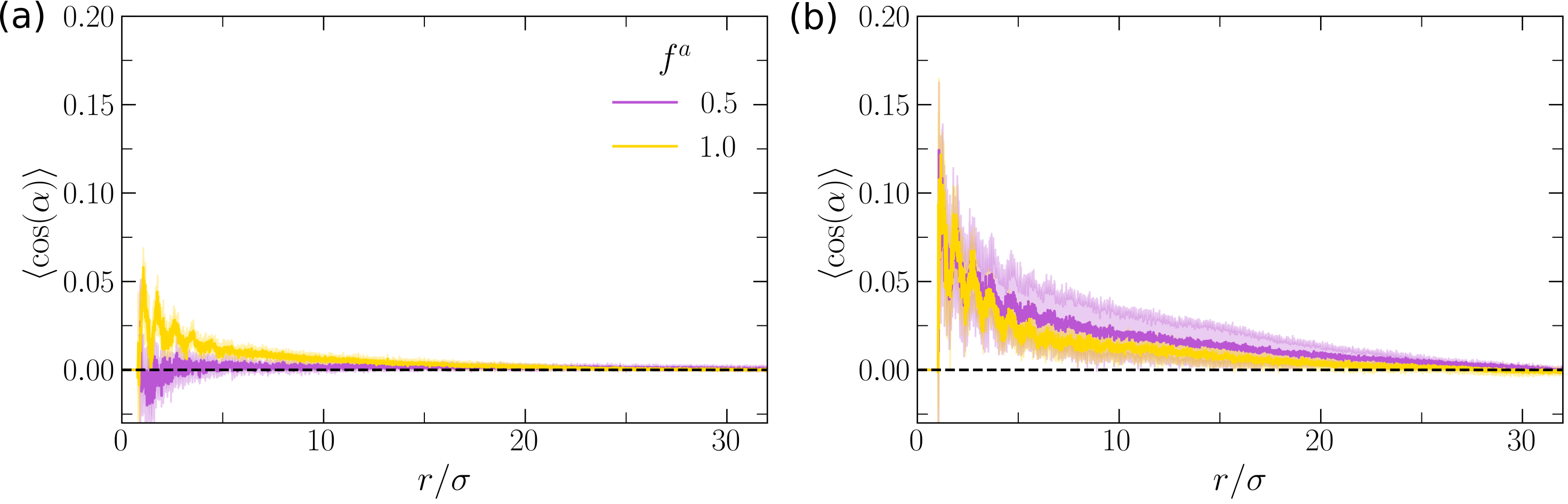}
    \caption{\textbf{Correlation of the polar order.} 
    (a-b)~The orientational polar order $\langle \cos(\alpha)\rangle$ for (a) $\kappa/k_bT=16$, $\rho\sigma^3=1.0$ and (b) $\kappa/k_bT=128$, $\rho\sigma^3=0.4$.  }
    \label{fig:Polar_Order}
\end{figure}

To investigate the length scale of the decay of polar order, we use the polar orientational correlation function $\big\langle\cos(\alpha)\big\rangle$. This is based on the same angle $\alpha$ as for the nematic orientational correlation function $g_{or}$ as the angle two neighboring bond vectors, which are a distance $r$ apart. Averaging over all bond vectors gives us the spatial length scale and strength of the polar order. Looking at two different activity levels in Sup. Fig. \ref{fig:Polar_Order}(a), we see a small magnitude of the polar order and a fast decay in the instability regime. Even for the stiffest polymers of $\kappa/k_BT=128$, we do not observe any significant polar order and in the nematic regime still decaying within $20 \sigma$, see Sup. Fig. \ref{fig:Polar_Order}(b). 

\section{III.~ Active semiflexible polymers in the dilute regime}\label{sec:Dilute_Phase}

To establish  the role of activity on conformational properties in the absence of crowding, we first investigate polymers in the dilute regime, $\rho\sigma^3\ll1$. Due to the GPU architecture of HOOMD-blue~\cite{Anderson2020}, it is computationally more efficient to simulate many polymers at very low density than to simulate a single polymer without inter-polymer interactions. We define the dilute regime as being below the overlap concentration $\rho\sigma^{3}\approx N^{-0.8}\approx6.25\cdot10^{-2}$~\cite{doi86}, such that the system effectively exhibits single-polymer behavior. In our simulations, we take a box of size $L_D=512\sigma$ with $M=1024$ polymers, giving an effective density of $\rho\sigma^{3}=2.44\times10^{-3}$.  

\begin{figure}[h]
    \centering
    \includegraphics[width=\linewidth]{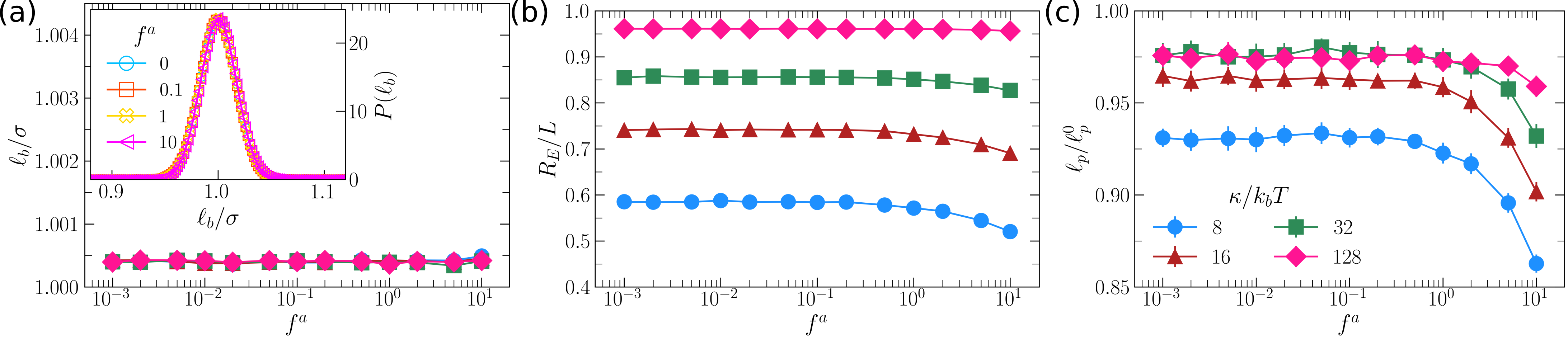}
    \caption{\textbf{Single-chain structural properties as a function of activity in the dilute limit. } 
    (a)~Effective persistence length $\ell_p$ with increasing activity.  
    (b)~Normalized end-to-end distance $R_e/L$ for different activity levels.  
    (c)~Average bond length $\ell_b$ as a function of activity. Inset: bond-length distribution at selected active forces.}
    \label{fig:Single_Chain_Activity_Structure}
\end{figure}

We first verify that t the bonds are not stretched due to activity and   the contour length $L=(N-1)\ell_b$ remains  constant as activity increases, as shown in Sup.~Fig.~\ref{fig:Single_Chain_Activity_Structure}(a) confirming that the average bond length remains constant up to $f^{a}\approx10$. Moreover, the distribution of bond lengths also remains narrow and unchanged with $f^a$. This confirms that we can treat the contour length as constant in all calculations. 

To distinguish collective effects from activity-induced changes in polymer structure, we examine the conformational properties  as  a function of activity level in the dilute limit. In Sup. Fig.~\ref{fig:Single_Chain_Activity_Structure}(b) and (c), both the end-to-end distance $R_e$ and the persistence length $\ell_p$ remain nearly constant up to an active force of $f^{a}\approx5.0$. For our collective simulations focusing on the activity range $f^{a}\leq2.0$,  hence any observed changes are due to collective effects arising from  interplay between activity and crowding rather than  activity-induced single-chain conformational changes.  

\section{IV.~ Geometric and effective tube radius}\label{sec:EffectiveTube}

\begin{figure}[h]
    \centering
    \includegraphics[width=0.8\linewidth]{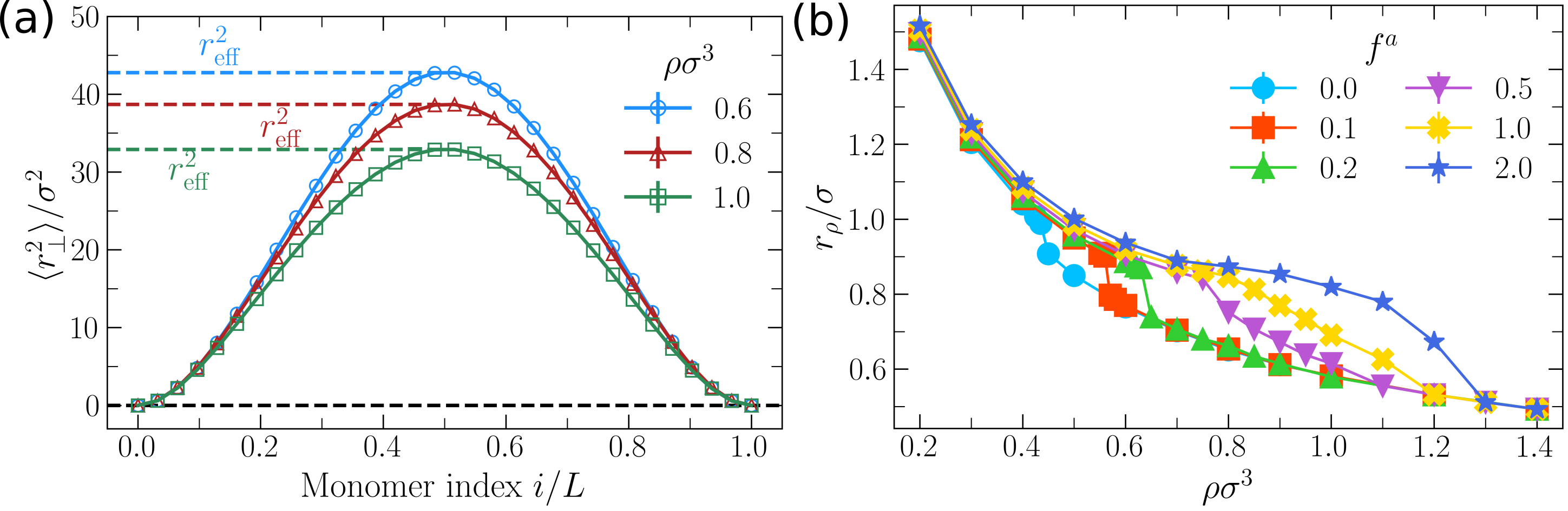}
    \caption{\textbf{Effect of density and activity on effective confinement.}  
    (a)~Distributions of the mean-square perpendicular distance $\langle r_{\perp}^2(i)\rangle$ along the polymer backbone at $f^{a}=2.0$ and increasing density. The maximum value defines the squared effective tube radius $r_{\text{eff}}^2$.  
    (b)~Geometric confinement radius from Eq.~\eqref{eqn:geometric} for $\kappa/k_BT=16$ at different activities and densities.}
    \label{smfig:r_eff}
\end{figure}

To describe the  motion of a single polymer in the high-density nematic phase, we adopt the tube picture description proposed in Ref.~\cite{Egorov2016}. In this picture, in a mean-field description, the effect of neighboring chains is to confine  a polymer  to an effective curvilinear cylindrical tube. From geometric arguments, one can estimate a density-induced tube radius $r_\rho$, based on the assumption that the density within a tube of length $\langle R_e^2\rangle^{1/2}$ and radius $r_\rho$ matches the overall density. This leads to
\begin{equation}\label{eqn:geometric}
   r_\rho=\sqrt{\frac{N}{\pi \rho \sqrt{\langle R_e^2\rangle}}}.
\end{equation}
As discussed in Ref.~\cite{Egorov2016} this geometric radius cannot capture the collective effects observed for passive semiflexible polymers.
Instead an effective tube radius $r_{\text{eff}}$ is defined based on measuring the transverse fluctuations of the polymer backbone relative to its end-to-end vector. For each monomer $i$, we first define the relative position vector 
\[
    \boldsymbol{r}_{i,\text{rel}}=\boldsymbol{r}_i-\boldsymbol{r}_1,
\]
where $\boldsymbol{r}_1$ is the position of the first monomer. Projecting $\boldsymbol{r}_{i,\text{rel}}$ onto the end-to-end vector $\vec{R}_e$ gives the parallel component,
\begin{equation}
    \boldsymbol{r}_{\parallel,i}=\frac{\boldsymbol{r}_{i,\text{rel}}\cdot\boldsymbol{R}_e}{{R}_e^2}\,\hat{\boldsymbol{R}}_e,
\end{equation}
where $\hat{\boldsymbol{R}}_e=\boldsymbol{R}_e/R_e$ is the unit end-to-end vector. The perpendicular displacement is then
\begin{equation}
     \langle r_{\perp}^2(i)\rangle = \Big\langle\big(\boldsymbol{r}_{i,\text{rel}}-\boldsymbol{r}_{\parallel,i}\big)\cdot\big(\boldsymbol{r}_{i,\text{rel}}-\boldsymbol{r}_{\parallel,i}\big)\Big\rangle,
\end{equation}
where $\langle .\rangle$ indicate the averaging over all the polymers in the system. We estimate the effective tube size as:
\[
    r_{\text{eff}}=\sqrt{\max_i \langle r_{\perp}^2(i)\rangle}.
\]
An example is shown in Sup.~Fig.~\ref{smfig:r_eff}(a) for $f^{a}=2.0$. Comparing $r_{\text{eff}}$ with the geometric radius $r_\rho$, we find that $r_{\text{eff}}$ is much larger, indicating that polymers are less constrained than expected from density-based considerations alone, see Sup.~Fig.~\ref{smfig:r_eff}(b).

\section{V.~ Simulations of single active polymers in a cylindrical tube }\label{sec:TubeDynamics}

To better investigate the effect of confinement, we simulated individual  polymers inside a cylindrical tube as a proxy for crowding-induced confinement. Following Ref.~\cite{MartinRoca2024}, we placed polymers with $N=32$ monomers in a box of size $L_x=L_y=2(R+\sigma)+\sigma$ and $L_z=128\sigma$. The box is periodic in the $z$-direction, and the walls at $R+\sigma$ interact with the polymer through a WCA potential, giving an effective tube radius $R$. Simulations were performed at a fixed bending stiffness of $\kappa/k_BT=16$, with WCA interactions between the monomers and activity values $f^{a}=0.0$--$2.0$.  

All polymers were initially aligned along the $z$-axis. The interaction strength $\varepsilon$ was gradually increased to $\varepsilon=1$, followed by equilibration for $50\tau$. Production runs were then carried out for $256$ independent polymers over $10^4\tau$, with data recorded every $100\tau$. Averages were computed over the last $9000\tau$ to obtain the mean end-to-end distance $R_e$. For each active force value, simulations were performed at confinement radii $R=1$--$16$, spanning the transition from shrunken to stretched conformations. The $R\to\infty$ limit was taken from the dilute simulations described in Sec.~\ref{sec:Dilute_Phase}.

\section{VI.~ The splay and bending deformation modes of the local nematic field}
\begin{figure}[h]
    \centering
    \includegraphics[width=0.4\linewidth]{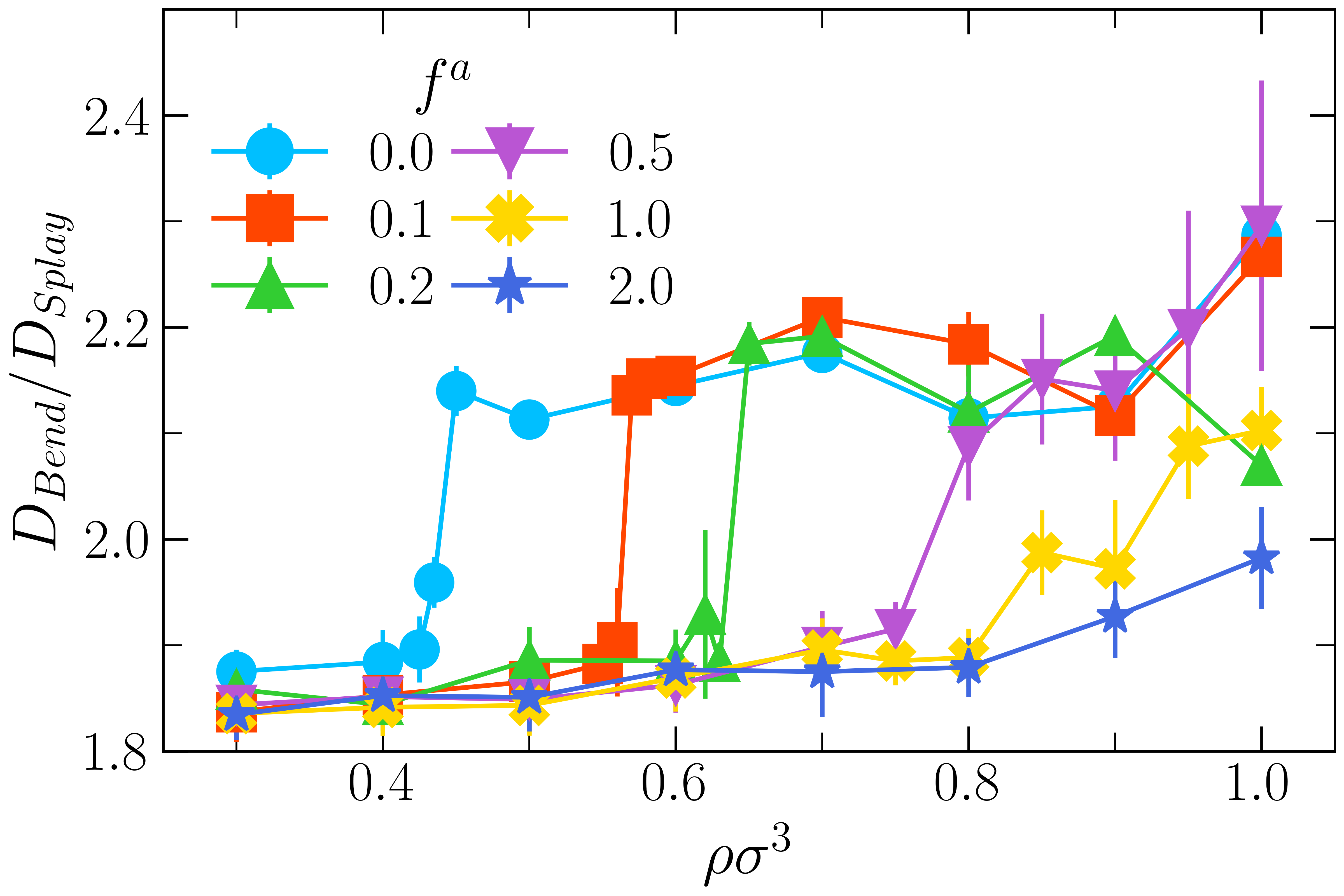}
    \caption{\textbf{Ratio of energy density of bend to splay deformations } 
    The ratio of the  mean  bending energy density $D_{Bend}$ to mean splay energy density   $D_{Splay}$ as a function  of density at different activity levels for bending stiffness $\kappa/k_BT=16$.}
    \label{fig:Bend_Splay}
\end{figure}

To quantify the energy density of  different modes of different deformation modes (bend and splay distortions), we first need to coarse grain our system to create a local nematic field. To define the local nematic field $\hat{\textbf{n}}(n_x,n_y,n_z)$, we made a grid of $64^3$ points with spacing $1\sigma$. For every grid point $(x_i,y_i,z_i)$, we make a selection of all the monomers closer than the cutoff distance $\xi= \big(\frac{81}{4\pi}\big)^{1/3}\sigma\approx1.86\sigma$, which gives a the total volume of $V_c=27\sigma^3$ per cell. These cell are allowed to overlap with each other, this is needed to create a smooth enough nematic field, which is needed to calculate the spatial derivatives later on. From the selected monomers, we use the associated bond vectors to calculate a local nematic tensor $\boldsymbol{Q}(n_x,n_y,n_z)$ using again Eq.~\ref{eq:Q_Nem_Local}. From the local nematic tensor $\boldsymbol{Q}(n_x,n_y,n_z)$, we find the nematic director field $\hat{\textbf{n}}(n_x,n_y,n_z)$ with associated nematic order parameter $S_\xi(n_x,n_y,n_z)$. Using the total selected monomers we find an additional density field $\rho_\xi(n_x,n_y,n_z)=N_c(n_x,n_y,n_z)/V_c$, $N_c$ is the number of bondvectors per grid cell. We then evaluate the bending deformation  field $\boldsymbol{d}_{Bend}=\big(\hat{\textbf{n}}\times(\nabla\times\hat{\textbf{n}})\big)$ and splay deformation field of $d_{splay}=\nabla \cdot\hat{\textbf{n}}$ from our local nematic order directors. These can be converted to their associated energy densities as $D_{Bend}=\rho_\xi S_\xi\lvert \boldsymbol{d}_{Bend}\rvert^2$ and $D_{Splay}=\rho_\xi S_\xi d_{splay}^2$, see \textit{Joshi et al.}\cite{Joshi2019} for details. To compare  the relative contribution of bend and splay   deformation modes, we compute the ratio of average strain energy in bend deformations to those in splay as $D_{Bend}/D_{Splay}=\frac{\langle D_{Bend}(n_x,n_y,n_z)\rangle}{\langle D_{Splay}(n_x,n_y,n_z)\rangle}$. In Sup. Fig. \ref{fig:Bend_Splay}, we plot $D_{Bend}/D_{Splay}$ against density for different activity levels.  We find that, concomitant with the increase of nematic order parameter, the mean  energy density of bending deformations increases relative to that of splay deformations. Hence, the onset of I--N transition in semiflexible polymers coincides with the predominance of bend over splay fluctuations. Interestingly, activity delays the predominance of bend deformation modes to higher densities, consistent with the activity-induced delay of the I--N transition.

\section{VII. Effect of flexibility degree on  the I--N transition of active polymers  }\label{sec:DiffKappa}

\begin{figure}[h]
    \centering
    \includegraphics[width=\linewidth]{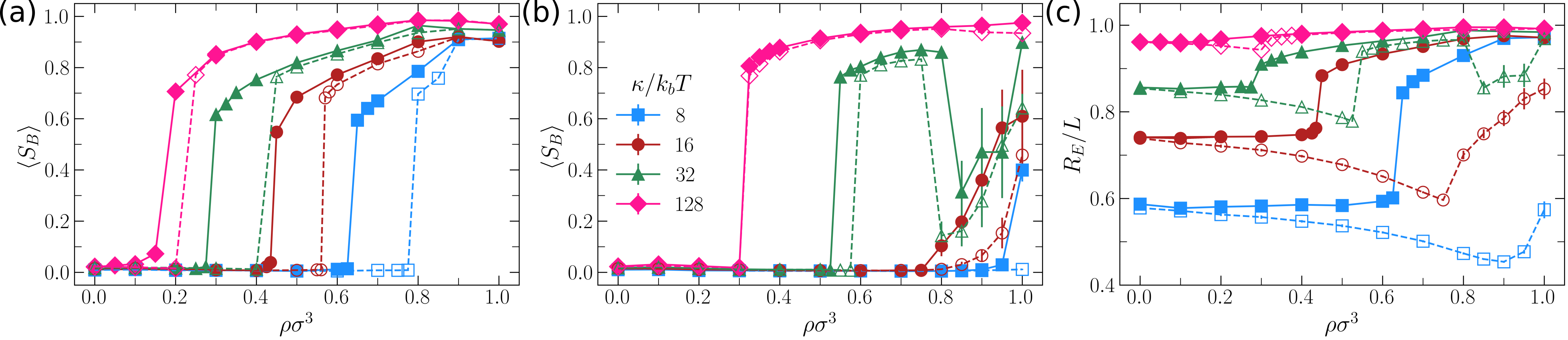}
    \caption{\textbf{The effect of the bending stiffness on nematic order}  
    (a-b) The time-averaged global nematic order parameter $\langle S_B \rangle$ is shown for varied values of bending stiffness $\kappa/k_bT=\{8,16,32,128\}$ of active polymers of contour length $L=31\, \sigma$ at different activity levels; (a) ~$f^{a}=0.0$ (filled symbols) and ~$f^{a}=0.1$ (open symbols), (b)~$f^{a}=0.5$ (filled symbols) and ~$f^{a}=1.0$ (open symbols). (c) The mean end-to-end distance $R_E$ normalized by counter length $L$ for the same bending stiffness for activities ~$f^{a}=0.0$ (filled symbols) and ~$f^{a}=0.5$ (open symbols).}
    \label{smfig:Kappa_Scaling}
\end{figure}

As discussed in the main text,  we investigated the effect of activity on the I--N transition of several bending stiffness values and found that they all display qualitatively similar behavior. In the main text, we  showed the I-N transition curves of  $\kappa/k_BT=16$. Here, we compare the results of $\langle S_B\rangle$ 
for all the  investigated  bending stiffness values $\kappa/k_BT=\{8,16,32,128\}$ at certain activity levels as shown in Sup.~Fig.~\ref{smfig:Kappa_Scaling}. 
In the passive limit,  the I--N  transition shifts to higher densities with increasing the degree of flexibility (lowering the bending stiffness). Hence, for the most flexible polymers  $\kappa/k_BT=8$,  the density of i I--N  transition is  the highest, consistent with the findings of Egorov \textit{et al.}~\cite{Egorov2016}, see Sup.~Fig.~\ref{smfig:Kappa_Scaling}(a). In the passive case ($f^{a}=0.0$) and at low activities ($f^{a}=0.1$), the I--N transition remains discontinuous for all the investigated $\kappa$ values.

However, at the higher activity of $f^{a}=0.5$, the nature of transition changes for $\kappa \le 32$ due to emergence of instability regime characterized by large temporal fluctuations of global nematic order parameter. Notably, for the bending stiffness of $\kappa/k_BT=32$,  at higher activity levels $f^a=0.5$ and 1, the system first displays a discontinuous I--N transition and it becomes only unstable   at higher densities, $\rho>0.75$ (see Sup.~Fig.~\ref{smfig:Kappa_Scaling}(b)). For the  stiffness case of $\kappa/k_BT=128$,  approaching the rod limit, the I--N transition remains continuous for all the investigated activity levels, see Sup.~Fig.~\ref{smfig:Kappa_Scaling}. For semiflexible polymers, with lower bending stiffness values, instabilities typically appeared for $f^{a}>0.5$. However, for $\kappa/k_BT=128$  the system remains completely stable up to $\rho\sigma^{3}=1.0$ and $f^{a}=2.0$. This shows that the I--N transition in semiflexible polymers results from a complex interplay of 
  activity, density their intrinsic degree of flexibility and is therefore not non-trivial. 

\section{VIII.~ Supplementary videos  }\label{sec:Video}
Supplementary Videos~1 and 2 show the dynamics of a single polymer with $N=32$ confined in a cylindrical tube of radius $R = 4\sigma$. Supplementary Video~1 corresponds to the passive case ($f^{a}=0$), whereas Supplementary Video~2 shows the same system at high activity ($f^{a}=2$).

Supplementary Video~3 displays the time evolution of a sample exhibiting stochastic transitions of the global nematic order parameter $S_{B}$ between   fully nematic  and  orientationally disordered states for a system at density $\rho\sigma^{3}=1.0$, activity $f^{a}=0.5$, and bending stiffness $\kappa/k_{B}T=16$. Polymers are colored according to their molecular ID.

Supplementary Video~4 shows the time evolution of a representative system in the unstable regime ($\rho\sigma^{3}=0.9$, $f^{a}=0.5$, $\kappa/k_{B}T=16$). Here, each polymer is colored by its local nematic order parameter $S^{m}_{\mathrm{loc}}$, as defined in Supplementary Sec.~\ref{sec:LocalNem}. The color scale is restricted to $0 \leq S^{m}_{\mathrm{loc}} \leq 1$.



\end{document}